\documentclass[a4paper,11pt]{article}
\usepackage[nohead, margin=1.5cm]{geometry}
\usepackage[utf8]{inputenc} 
\usepackage[T1]{fontenc}
\usepackage{appendix}
\usepackage{enumitem}
\usepackage{dsfont}
\usepackage{amsmath, esint}
\usepackage{amsfonts}
\usepackage{amssymb}
\usepackage{bbold}
\usepackage{graphicx}%
\usepackage[small,bf]{caption}
\usepackage{subfig}
\usepackage{float}
\usepackage{xcolor}
\usepackage{multirow}
\usepackage[
colorlinks=true]{hyperref}
\usepackage{slashed}
\usepackage{array}
\usepackage{epsfig} 
\usepackage{feynmf}
\usepackage[normalem]{ulem}
\usepackage{comment}
\usepackage{soul}
\usepackage{cancel}

\usepackage{color}
\usepackage{listings}
\definecolor{darkgreen}{rgb}{0,0.35,0}
\definecolor{Rood}{rgb}{1, 0, 0}

\newcommand\redsout{\bgroup\markoverwith{\textcolor{red}{\rule[0.4ex]{3pt}{0.7pt}}}\ULon}

\providecommand{\U}[1]{\protect\rule{.1in}{.1in}}

\newcommand{\bepsilon}{\bar{\epsilon}}

\newcommand{\btheta}{\bar{\theta}}
\newcommand{\bchi}{\bar{\chi}}
\newcommand{\blambda}{\bar{\lambda}}
\newcommand{\bpsi}{\bar{\psi}}
\newcommand{\bsigma}{\bar{\sigma}}

\newcommand{\bXi}{\bar{\Xi}}
\newcommand{\dalpha}{\dot{\alpha}}

\newcommand{\dbeta}{\dot{\beta}}

\newcommand{\p}{\partial}
\newcommand{\Tr}{\ensuremath{\mathrm{Tr}}}

\begin{document}

\title{\textbf{
Revisiting the Abelian $N=1$ super Stueckelberg model 
}}
\date{}
	
\author{
\textbf{M.~A.~L.~Capri}$^{a}$\thanks{caprimarcio@gmail.com}\,\,,
\textbf{D.~R.~Granado}$^{b}$\thanks{diegorochagranado@gmail.com}\,\,,
\textbf{I.~F.~Justo}$^{c}$\thanks{igorfjusto@fat.uerj.br}\,\,,
\textbf{L.~S.~S.~Mendes}$^{a}$\thanks{santiago.petropolis@outlook.com}\,\,,\\[2mm]
{\small \textnormal{$^{a}$  
\it Departamento de F\'{\i }sica Te\'{o}rica, Instituto de F\'{\i }sica,}} \\
{\small \textnormal{\phantom{$^{a}$} UERJ - Universidade do Estado do Rio de Janeiro,}}\\
{\small \textnormal{ \it Rua S\~{a}o Francisco Xavier 524, 20550-013 Maracan\~{a}, Rio de Janeiro, Brasil}\normalsize}
\\ 
{\small \textnormal{$^{b}$  
\it CBPF - Centro Brasileiro de Pesquisas F\'isicas,}} \\
{\small \textnormal{ \it Rua Doutor Xavier Sigaud 150, 22290-180, Rio de Janeiro, Brazil}\normalsize}
\\ 
{\small \textnormal{$^{c}$ 
\it Departamento de Matemática, Física e Computação, Faculdade de Tecnologia,}}\\
{\small \textnormal{\phantom{$^{c}$} UERJ - Universidade do Estado do Rio de Janeiro,}}\\
{\small \textnormal{\phantom{$^{c}$} 
\it Av. Dr. Omar Dibo Calixto Afrange, s/n, Resende - RJ, Brasil}\normalsize}
}

\maketitle

\begin{abstract}
The Abelian super Stueckelberg model (ASSM) in the Wess--Zumino (WZ) gauge is revisited, and the actual set of supersymmetric (SUSY) transformation is derived. In particular, we verified that the SUSY transformation of the super Stueckelberg sector compensates the gauge fixing condition imposed on the vector superfield, leading to a mix between the field components of both sectors. We also discuss the possibility to construct an extension of the ASSM with infinite self interacting terms.
\end{abstract}

\section{Introduction}

Originally constructed in the Abelian gauge theory, the Stueckelberg model (SM) describes a massive gauge field while preserving gauge invariance by introducing a compensating scalar field, \cite{Stueckelberg:1938hvi}. The investigation of the Abelian massive gauge fields began with the construction of gauge invariant descriptions of massive vector bosons, as a massive photon of the hypercharge in the electroweak theory. In recent years, renewed interest in the SM has arisen in cosmological scenarios where massive photons would play a role, \cite{Kouwn:2015cdw,Lemos:2025qyh}. Since its first proposal, the SM has been frequently adopted as a gauge invariant alternative to the Higgs mechanism of mass generation, \cite{Burnel:1986kb,Ruegg:2003ps}, despite its issues with the conflict between renormalizability and unitarity in Yang-Mills (YM) theories ({\it cf.} \cite{Kunimasa:1967zza} for the first YM SM formulation.). The Abelian formulation of the SM has been proved to be perturbatively renormalizable and unitary, which is not the case in the YM scenario, \cite{Lowenstein:1972pr,Delbourgo:1987np,vanHees:2003dk} ({\it c.f.} \cite{Ruegg:2003ps} for a detailed review of the SM and for discussions on renormalizability and unitarity). In \cite{Delbourgo:1987np,Delbourgo:1986wz} the authors propose a variant of the Stueckelberg model in YM theories where they discuss the power-counting renormalizability and unitarity in the Landau gauge. An all-order renormalizable Stueckelberg-like model, with the transversality constraint, is developed in 2016 by the authors of \cite{Capri:2016ovw}, for the class of Linear Covariant gauges. This approach has recently been developed to address the fate of Gribov copies in a BRST invariant approach, \cite{Capri:2015ixa,Capri:2016aqq,Capri:2017bfd}.

Over the years, the Abelian supersymmetric Stueckelberg model (ASSM) has been proposed and proved to be power-counting renormalizable, \cite{Delbourgo:1975uf}. The authors of \cite{Delbourgo:1975uf} pointed to the role played by the scalar superfield (which is an object of superspace presented in the Appendix \ref{appendix_A}) that cancels the ultraviolet longitudinal contributions of the vector super field propagator in covariant gauges, which is technically necessary to ensure unitarity to the ASSM. Such supersymmetric (SUSY) extension leads to the existence of massive bosons and fermions degrees of freedom with manifest gauge invariance. Since 1998 the ASSM is a subject of textbook \cite{Buchbinder:1998twe}, and has been investigated in a multitude of scenarios. In \cite{Kors:2004ri,Kors:2004iz,Kors:2004dx,Kors:2005uz} the authors propose a Stueckelberg extension of the minimal SUSY Standard Model with one vector and one chiral superfields, opposed to the Abelian extension with a Higgs scalar superfield. Another example can be found in \cite{Ghaffari:2018lmo,Kuzmin:2002nt,Kuzmin:2002xxv}, where spontaneous symmetry breaking of supersymmetry is investigated in the ASSM in the presence of a Fayet-Iliopoulos $D$-term. Furthermore, in \cite{Vinze:2020vfh} the authors studied the infrared divergences in the massless limit of a $N=1$ SUSY massive $U(1)$ gauge theory coupled to matter and the Stueckelberg superfields. The ASSM has also been explored in the context of dark matter candidates \cite{Coriano:2020esy, Han:2020oet}. The non-Abelian formulation of the super Stueckelberg model has been the subject of recent investigations, \cite{Nishino:2013kxa,Nishino:2013oea}. In \cite{Capri:2018gpu} the authors investigated the renormalizability of a super Yang-Mills (SYM) action in the presence of a generalized Stueckelberg-like gauge invariant mass term, constructed by means of a power series of a gauge invariant and transverse superfield \footnote{The denomination ``Stueckelberg-like'' comes from the constraint of transversality of the gauge invariant superfield, which is not present in the original formulation of the Stueckelberg model but seems to be crucial for the renormalizability in the non-Abelian case \cite{Capri:2016ovw}.}. In our present work, we investigate the SUSY set of transformations, in the (3+1) Minkowsky spacetime, of the Abelian version of the model proposed in \cite{Capri:2018gpu}.

The Algebraic Renormalization (AR) procedure is a very powerful approach to investigate the ultraviolet (UV) stability of a model at all order in perturbation theory, with the set of symmetries of the model playing a central role to control the UV divergencies, \cite{Piguet:1995er}. The starting point to probe the renormalizability at all order is to collect the greatest number of symmetries of the classical action in a specific gauge, and verify if such symmetries can be promoted to quantum symmetries of the effective action at any order of quantum correction. In SYM theories in the Wess-Zumino (WZ) gauge, \cite{Wess:1974jb}, for instance, one of the central symmetries in the AR procedure is the set of nilpotent super BRST transformation, where each field component transforms as a BRST plus supersymmetry transformations ({\it cf.} \cite{Maggiore:1994dw,Maggiore:1995gr,Capri:2014jqa,Capri:2018lru}, and also in the SUSY electroweak theory \cite{Hollik:2002mv}). The existence of such nilpotent symmetry rely on the algebra of supersymmetry that in the WZ gauge closes on translations plus gauge transformation,
\begin{align}
\left[ \delta_2 \,, \delta_1 \right] \varphi(x) ~=~ 
\left( i{\epsilon_1}{\sigma}^{\mu}\bepsilon_2  - i{\epsilon_2}{\sigma}^{\mu}\bepsilon_1 \right) \p_{\mu} \varphi(x) 
+ \delta_{\rm gauge}(\omega)\varphi(x)\,,
\label{susy_algebra}
\end{align}
where $\varphi(x)$ stands for any field component of the superfield, $\epsilon_{1}$ and $\epsilon_{2}$ are two component fermionic infinitesimal parameters of supersymmetry transformations $\delta_{1}$ and $\delta_{2}$, respectively ({\it cf.} \cite{Gates:1983nr,Buchbinder:1998twe,Piguet:1995er}), and the gauge transformation parameter $\omega$ depends on the gauge field $A_\mu (x)$,
\begin{align}
\omega ~=~ \left( i{\epsilon_1}{\sigma}^{\mu}\bepsilon_2 - i{\epsilon_2}{\sigma}^{\mu}\bepsilon_1 \right)A_{\mu}(x)\,.
\end{align}

In this paper we propose to revisit the Abelian $N=1$ super Stueckelberg-like model in the WZ gauge, starting with the most general gauge invariant mass term in superspace, first introduced in \cite{Capri:2018gpu}, which is a power series of\footnote{The notation used  here is slightly different from \cite{Capri:2018gpu}. The correspondence are given by $V^{H}\equiv V^{S}$, $\Xi\equiv S$ and $\bXi\equiv \bar S$, where the l.h.s are the notations of \cite{Capri:2018gpu}.} 
\begin{align}
V^{S} &~=~ V + i\left( S - \bar{S} \right)\,,
\end{align}
with $S$$(\bar S)$ being a chiral (anti-chiral) superfield accounting for the super Stueckelberg sector, and $V$ standing for the vector superfield. As will become explicit later, $V^{S}$ is a dimensionless gauge invariant vector superfield, then the ASSM can be extented by contributions of $(V^{S})^{n}$ with $n > 2$, introducing self interacting terms of the field components of $V$ and \( i\left( S - \bar S \right) \), which may bring interesting non-trivial behavior in the UV regime, even in the Abelian case. Starting with the well known ASSM, \cite{Buchbinder:1998twe}, with mass term proportional to $V^{S}V^{S}$, the existence of higher order terms of self interact field components are controlled by Ward Identities. Then the UV regime of the extended version of ASSM, with extra powers of $V^{S}$, must be investigated. However, in order to investigate the set of SUSY transformations in the WZ gauge, it is enough to consider only the quadratic contribution of $V^S$.

This paper is organized as follows. The Section \ref{Abelian Stueckelberg-Like model} is devoted to carefully set up the ASSM, starting with the most general gauge invariant mass term of \cite{Capri:2018gpu} in superspace, and then derive its expression in the Minkowski (3+1) dimension spacetime. Besides, still in Section \ref{Abelian Stueckelberg-Like model} a detailed review of the gauge and supersymmetry transformations is performed. The Section \ref{The Wess-Zumino gauge model} is devoted to the Wess-Zumino gauge, imposed to the ASSM with contributions of $V^{S}V^{S}$, and a brief appointment is made on the set of supersymmetry of the model. In the Section \ref{section4} we present a new set of transformations that is an explicit supersymmetry of the model, that closes on the algebra \eqref{susy_algebra}. In Appendix \ref{appendix_A} our notations, conventions and definitions are available, and in Appendix \ref{appendix_B} we have a brief discussion on the construction of the ASSM independent of the compensating fields without the necessity of employing the WZ gauge.

\section{The super Abelian Stueckelberg-like model}
\label{Abelian Stueckelberg-Like model}

In this section, the extended Abelian super Stueckelberg model (EASSM) is set up, following the non-Abelian model studied in \cite{Capri:2018gpu}. Our notations and conventions are mostly the same as those of \cite{Capri:2018gpu}, but further details can be found in Appendix \ref{appendix_A}.

\subsection{The model in superspace}
\label{subsec2.1}
	
Let us start by defining the superfields in terms of their components. The vector superfield reads
\begin{align}
\label{V}
V &~=~ 
C 
+ i\theta \chi 
- i \btheta\bchi 
+ \theta \sigma^{\mu}\btheta A_{\mu} 
+ \frac{i}{2}\theta^{2}M 
- \frac{i}{2}\btheta^{2}\bar{M} 
+ i\theta^{2}\btheta\left( \blambda + \frac{i}{2}\bsigma^{\mu}\p_{\mu}\chi \right) \nonumber \\
& \phantom{~=~}
- i\btheta^{2}\theta\left( \lambda + \frac{i}{2}\sigma^{\mu}\p_{\mu}\bchi \right)  
+ \frac12 \theta^{2}\btheta^{2} \left( D - \frac12\p^{2}C \right)\,.
\end{align}
Here, the components $(C,\chi,\bar{\chi},M,\bar{M})$ are recognized as the compensating fields, the vectorial component $A_{\mu}$ is the gauge field, $(\lambda,\bar{\lambda})$ are the gaugino fields and $D$ is an auxiliary field.
The Stueckelberg sector, $i\left( S - \bar S \right)$, is written in terms of the chiral and anti-chiral superfields\footnote{See Appendix \ref{appendix_A} for definitions of chiral superfields.}, namely
\begin{equation}
\begin{aligned}
\label{chiral_sufields}
S &~=~ 
(\rho + ia) 
+ \sqrt{2} \theta \psi 
+ i\theta\sigma^{\mu} \btheta\p_{\mu}(\rho + ia) 
-  \theta^2 f 
+ \frac{i}{\sqrt{2}} \theta^2 \btheta\bsigma^{\mu}  \p_{\mu}\psi
- \frac{1}{4} \theta^2 \btheta^2 \p^{2}(\rho + ia)\,, 
\\
\bar S &~=~ 
(\rho - ia) 
+ \sqrt{2} \btheta \bpsi 
- i\theta\sigma^{\mu} \btheta\p_{\mu}(\rho - ia) 
-  \btheta^{2} \bar{f} 
+ \frac{i}{\sqrt{2}} \btheta^2 \theta \sigma^{\mu}\p_{\mu}\bpsi 
- \frac{1}{4} \btheta^2 \theta^2 \p^{2}(\rho - ia)\,.
\end{aligned}
\end{equation}
The gauge invariant superfield $V^{S}$ can be constructed as
\begin{align}
\label{VH}
V^{S} &~=~ V + i\left( S - \bar S \right) \nonumber \\
&~=~ 
\big[C - 2a\big] 
+ i\theta \big[ \chi + \sqrt{2}\psi \big] 
- i\btheta\big[ \bchi + \sqrt{2}\bpsi \big] 
+ \theta\sigma^{\mu}\btheta \big[ A_{\mu} - 2\p_{\mu}\rho \big] 
+ \frac{i}{2} \theta^{2}\left[ M - 2f \right]
\nonumber \\
& \phantom{~=~}
- \frac{i}{2}\btheta^{2}\left[ \bar{M} - 2\bar{f} \right] 
+ i\theta^{2}\btheta \left[ \blambda + \frac{i}{2}\bsigma^{\mu}\p_{\mu} \left( \chi + \sqrt{2}\psi \right) \right] 
- i\btheta^{2}\theta \left[ \lambda + \frac{i}{2} \sigma^{\mu}\p_{\mu}\left( \bchi + \sqrt{2}\bpsi\right) \right]
\nonumber \\
& \phantom{~=~} 
+ \frac12 \theta^{2}\btheta^{2} \left[ D - \frac12\p^{2}(C -2a) \right]\,.
\end{align}
We can also define a shorter notation
\begin{align}
\label{VS}
V^{S} &~=~ 
C^s 
+ i\theta \chi^s 
- i \btheta\bchi^s 
+ \theta \sigma^{\mu}\btheta A_{\mu}^s 
+ \frac{i}{2}\theta^{2}M^s
- \frac{i}{2}\btheta^{2}\bar{M}^s
+ i\theta^{2}\btheta\left( \blambda + \frac{i}{2}\bsigma^{\mu}\p_{\mu}\chi^s \right) \nonumber \\
& \phantom{~=~}
- i\btheta^{2}\theta\left( \lambda + \frac{i}{2}\sigma^{\mu}\p_{\mu}\bchi^s \right)  
+ \frac12 \theta^{2}\btheta^{2} \left( D - \frac12\p^{2}C^s \right)\,,
\end{align}
where
\begin{equation}
\begin{aligned}
\label{hfields}
C^{s} ~=~ (C-2a)\,, 
\qquad \chi^{s} ~=~ (\chi + \sqrt{2}\psi)\,, 
\qquad \bchi^{s} ~=~ (\bchi + \sqrt{2}\bpsi)\,,
\\
A^{s}_{\mu} ~=~ (A_{\mu} - 2\p_{\mu}\rho)\,,  
\quad 
M^{s} ~=~ M - 2f\,, 
\quad 
\bar{M}^{s} ~=~ \bar{M} - 2\bar{f}\,.
\end{aligned}
\end{equation}
The gauge invariance of $V^{S}$ is obtained by demanding that
\begin{equation}
\label{gauge_transf_relations}
\begin{aligned}
\partial_{\mu}\delta\rho=\frac{1}{2}\delta A_{\mu}\,,\qquad
\delta a=\frac{1}{2} \delta C\,,\qquad
\delta \psi=\frac{1}{\sqrt{2}}\delta\chi\,,\\
\delta \bar\psi=\frac{1}{\sqrt{2}}\delta\bar\chi\,,\qquad
\delta f=\frac{1}{2}\delta M\,,\qquad \delta \bar f=\frac{1}{2}\delta \bar M\,,
\end{aligned}
\end{equation}
with $\delta$ standing by the gauge variations. The gauge variations of the components of the vector superfield are obtained from \eqref{vec_field_gtransf}, as we shall see later.

Therefore, the action of EASSM in superspace reads
\begin{align}
\label{susy_act}
\mathcal{S}_{\rm EASSM} &~=~ 
\frac{1}{64}\int\, d^{4}x\, d^{2}\theta \ \ {W}(V){W}(V) + {c.c.} 
+ m^{2} \int\, d^{4}x\, d^{2}\theta \, d^{2}\btheta \ \ {\cal M}(V^{S})\,,
\end{align}
where, 
\begin{align}
W_\alpha (V) &~=~ 
- \bar{D}\bar{D}D_{\alpha}V\, \\
\overline{W}^{\dot{\alpha}} (V) &~=~ 
- DD\bar{D}^{\alpha}V\,,\nonumber
\end{align}
and 
\begin{align}
\label{mass_superspace}
{\cal M}(V^{S}) &~=~
V^{S}V^{S} 
+ \alpha_1\,V^{S}V^{S}V^{S} 
+ \alpha_{2}\,V^{S}V^{S}V^{S}V^{S} 
+ \cdots
\end{align}
The $\alpha_{i}$ (with $i ~=~ 1,\,2,\,3,\,\dots$) are dimensionless coupling constants. It must be clear that the action \eqref{susy_act} is obtained as the particular Abelian limit case of that one proposed in \cite{Capri:2018gpu}. There the $V^{S}$ superfield is imposed to be transversal, which is essential to demonstrate its all order renormalizability. Here, instead, in the EASSM we are not imposing such transversality constraint on $V^{S}$, which makes the UV stability of this model still open to investigations.

\subsection{The model in Minkowski spacetime}
\label{subsec2.2}

By integrating out the Grassmann valued coordinates $(\theta,\btheta)$ form \eqref{susy_act}, the field component expression of \eqref{susy_act}, in Minkowski spacetime, can be derived.  As is well known, the dynamical contribution, from ${W}(V){W}(V)$, becomes,
\begin{align}
\frac{1}{64}\int\, d^{4}x\, d^{2}\theta \ \ {W}(V){W}(V) + {c.c.} &~=~
\int\, d^{4}x \ \left\{
-\frac{1}{4} f^{\mu\nu} f_{\mu\nu} 
- i \lambda \sigma^{\mu} \p_{\mu}\bar{\lambda}
+ \frac12 D^{2}
\right\}\,,
\label{abelian_act}
\end{align}
with $f_{\mu\nu} = \p_{\mu}A_{\nu} - \p_{\nu}A_{\mu}$. It is remarkable that in the Abelian case the compensating fields completely disappear from the action when the Grassmann  coordinate integrations are performed, independently of the imposition of the WZ gauge. The the mass term contributuion, coming from \eqref{mass_superspace}, reads, until the fourth order,
\begin{align}
\label{mass}
& \int\, d^{4}x\, d^{2}\theta \, d^{2}\btheta \ \ {\cal M}(V^{S})  
\nonumber \\
&~=~
\int\, d^{4}x\, \left[\frac{1}{2}A^{s}{}^{\mu}A_{\mu}^{s} 
- \bchi^{s}\left(\blambda + \frac{i}{2}\bsigma^{\mu}\p_{\mu}\chi^{s}\right)
- \chi^{s}\left(\lambda + \frac{i}{2}\sigma^{\mu}\p_{\mu}\bar{\chi}^{s}\right)
+ \left(D - \frac{1}{2}\p^{2}C^{s}\right)C^{s}
\right.
\nonumber \\
& \phantom{~=~}
\left.
+\frac{1}{2}\bar{M}^{s}M^{s}\right] 
+ \alpha_{1} \int\, d^{4}x\, \left[
\frac{3}{2}A^{s}{}^{\mu}A_{\mu}^{s}C^{s}
- 3\bar{\chi}^{s}\left(\blambda + \frac{i}{2}\bsigma^{\mu}\p_{\mu}\chi^{s}\right)C^{s} 
- 3\chi^{s}\left(\lambda + \frac{i}{2}\sigma^{\mu}\p_{\mu}\bchi^{s}\right)C^{s}
\right.
\nonumber \\
& \phantom{~=~}
\left.
+ \frac{3}{2}\left(D - \frac{1}{2}\p^{2}C^{s}\right)(C^{s})^{2}
+ \frac{3}{2}\bar{M}^{s}M^{s}C^{s} 
+ \frac{3i}{4}\bchi^{s}\bchi^{s}M^{s} 
- \frac{3i}{4} \chi^{s}\chi^{s}\bar{M}^{s}
+ \frac{3}{2}\chi^{s}\sigma^{\mu}\bchi^{s}A^{s}_{\mu}
\right]
\nonumber \\
& \phantom{~=~} 
+ \alpha_{2} \int\, d^{4}x\, \left[
\frac{6}{2}{A^{s}}^{\mu}A^{s}_{\mu}C^{s}C^{s} 
- 6\bchi^{s}\left(\blambda + \frac{i}{2}\bar{\sigma}^{\mu}\p_{\mu}\chi^{s}\right)(C^{s})^{2}
- 6\chi^{s}\left(\lambda + \frac{i}{2} \sigma^{\mu}\p_{\mu}\bar{\chi}^{s}\right)(C^{s})^{2}
\right.
\nonumber \\
& \phantom{~=~}
+ \frac{4}{2}\left(D - \frac{1}{2}\p^{2}C^{s}\right)(C^{s})^{3}
+ 3\bar{M}^{s}M^{s}(C^{s})^{2}
+ 3i \bchi^{s}\bchi^{s}M^{s}C^{s}
- 3i\chi^{s}\chi^{s}\bar{M}^{s}C^{s}  
\nonumber \\
& \phantom{~=~}
\left.
+ 6\chi^{s}\sigma^{\mu}\bchi^{S}A^{s}_{\mu} C^{s}
+ \frac{3}{2}\bchi^{s}\bchi^{s}\chi^{s} \chi^{s} \right] 
+ \mathcal{O}(5)\,.
\end{align}
The difference between the usual ASSM, \cite{Buchbinder:1998twe}, and its extended version is evident from \eqref{mass}. The existence of infinite $\alpha_{i}$ terms is formally allowed due to the dimensionless character of $C^{s}$. Notice, for instance, the presence of new vertices, such as the $\chi^{s}\sigma^{\mu}\bchi^{s}A^{s}_{\mu}$, and that they always appear in the next $\alpha_{i}$ term multiplied by powers of $C^{s}$. Furthermore, contributions for $\alpha_{i}$, with $i>2$, brings no new vertices (but only those from $\alpha_{1}$ and $\alpha_{2}$ times powers of $C^{s}$), so that the dynamic of $C^{s}$ is of great importance to understand the physical meaning of these higher order contributions.

To our presente propose, it is enough to be restricted to the ASSM, considering only the bilinear terms of \eqref{mass}, so that our starting action is
\begin{align}
\label{action2}
\mathcal{S} &=~
\int\, d^{4}x \ \left\{
-\frac{1}{4} f^{\mu\nu} f_{\mu\nu} 
- i \lambda \sigma^{\mu} \p_{\mu}\bar{\lambda}
+ \frac12 D^{2}
\right\}
+ m^2\int\, d^{4}x\, d^{2}\theta \, d^{2}\btheta \ \ V^{S}V^{S} \,,
\end{align}
with
\begin{align}
\label{mass_2}
& \int\, d^{4}x\, d^{2}\theta \, d^{2}\btheta \ V^{S}V^{S}  ~=~ 
\int\, d^{4}x\, \left[\frac{1}{2}A^{S}{}^{\mu}A_{\mu}^{S} 
- \bchi^{S}\left(\blambda + \frac{i}{2}\bsigma^{\mu}\p_{\mu}\chi^{S}\right)
- \chi^{S}\left(\lambda + \frac{i}{2}\sigma^{\mu}\p_{\mu}\bar{\chi}^{S}\right)
\right.
\nonumber \\
& \phantom{\int\, d^{4}x\, d^{2}\theta \, d^{2}\btheta \ V^{S}V^{S}  ~=~}
\left.
+ \left(D - \frac{1}{2}\p^{2}C^{S}\right)C^{S} 
+\frac{1}{2}\bar{M}^{S}M^{S}\right] \,.
\end{align}
Notice that the mass potential term \eqref{mass_2} makes the ASSM \eqref{action2} dependent on the compensating fields $(C,\, \chi,\, \bchi,\, M,\, \bar{M})$, so that the WZ gauge fixing procedure must be carefully carried out.

\subsection{Gauge invariance}

In superspace, the Abelian gauge transformation of the vector superfield $V$ reads
\begin{align}
\label{g.transf.1}
V ~\to~ V' &=~ V +i(\Lambda - \bar{\Lambda})\,,
\end{align}
where $\Lambda$ and $\bar{\Lambda}$ stand, respectively, for the chiral and anti-chiral superfields responsible for the gauge transformation. The gauge transformations of the chiral and anti-chiral superfields $(S,\bar{S})$ are
\begin{equation}
\begin{aligned}
\label{g.transf.2}
S~&\to~S'=S-\Lambda\,,\\
\bar{S}~&\to~\bar{S}'=\bar{S} - \bar\Lambda\,.
\end{aligned}
\end{equation}
As a consequence of \eqref{g.transf.1} and \eqref{g.transf.2},  $V^S$, presented in \eqref{VH}, is left invariant:
\begin{align}
\label{VH.gauge.inv}
V^{S} ~\to~ V^{S} ~=~ {V^S}' + i\left( S' - \bar{S}' \right) ~=~ V^{S}\,.
\end{align}
Besides, since $\Lambda$ and $\bar{\Lambda}$ are chiral and anti-chiral superfields, their field component expression reads,
\begin{equation}
\begin{aligned}
\label{gauge_tranf_param}
\Lambda &~=~ 
z 
+ \sqrt{2}\theta\eta 
+ i\theta\sigma^{\mu}\btheta \p_{\mu}z 
- \theta^{2} h 
- \frac{i}{\sqrt{2}}\theta^{2}\p_{\mu}\eta\sigma^{\mu}\btheta
- \frac{1}{4}\theta^{2}\btheta^{2}\p^{2}z\,,
\\
\bar{\Lambda} &~=~ 
\bar{z} 
+ \sqrt{2}\btheta\bar{\eta} 
- i\theta\sigma^{\mu}\btheta \p_{\mu}\bar{z} 
- \btheta^{2} \bar{h} 
+ \frac{i}{\sqrt{2}}\btheta^{2}\theta\sigma^{\mu}\p_{\mu}\bar{\eta}
- \frac{1}{4}\theta^{2}\btheta^{2}\p^{2}\bar{z}\,,
\end{aligned}
\end{equation}
so that the field component expression of the gauge transformed $V'$ (according to \eqref{g.transf.1}) becomes,
\begin{align}
\label{gtransf_sV}
V' &~=~ 
\left[ C - 2 {\rm Im}(z) \right] 
+ i\theta \left[\chi + \sqrt{2}\eta \right] 
- i \btheta\big[\bchi + \sqrt{2}\bar{\eta}\big] 
+ \theta \sigma^{\mu}\btheta \left[A_{\mu} -2\p_{\mu}{\rm Re}(z)\right] 
+ \frac{i}{2}\theta^{2}\left[M - 2h\right]
\nonumber \\
& \phantom{~=~} 
- \frac{i}{2}\btheta^{2}\left[\bar{M} - 2 \bar{h}\right] 
+ i\theta^{2}\btheta\left[ \blambda + \frac{i}{2}\bsigma^{\mu}\p_{\mu}\left(\chi + \sqrt{2} \eta \right) \right]
- i\btheta^{2}\theta\left[ \lambda + \frac{i}{2} \sigma^{\mu} \p_{\mu}\left(\bchi + \sqrt{2}\bar{\eta} \right) \right]
\nonumber\\
& \phantom{~=~} 
+ \frac12 \theta^{2}\btheta^{2} \left[ D - \frac12\p^{2}\left( C - 2 {\rm Im}(z) \right) \right]\,.
\end{align}
From \eqref{gtransf_sV} the Abelian gauge transformation of each field component can be read off,
\begin{equation}
\begin{aligned}
\label{vec_field_gtransf}
C' ~=~ C -2Im(z), 
&\qquad 
\chi' ~=~ \chi + \sqrt{2}\eta\,,\qquad 
\bchi' ~=~ \bchi + \sqrt{2}\bar{\eta}\,,\\ 
M' ~=~ M-2h\,,&\qquad
\bar{M}' ~=~ \bar{M} - 2 \bar{h}\,,\qquad
A'_{\mu} ~=~ A_{\mu} - 2 \p_{\mu}Re(z)\,,\\
\lambda' ~=~ \lambda\,,&\qquad
\blambda' ~=~ \blambda\,,\qquad
D' ~=~ D\,.
\end{aligned}
\end{equation}
Likewise, the set of gauge transformations of each field component of the (anti-)chiral superfields can be derived from \eqref{g.transf.2} and \eqref{gauge_tranf_param}. Namely,
\begin{equation}
\begin{aligned}
\label{chi_field_gtransf}
\rho' ~=~ \rho - Re(z)\,, 
&\qquad 
a' ~=~ a - Im(z)\,,
\\
\psi' ~=~ \psi - {\eta}\,, 
&\qquad 
\bpsi' ~=~ \bpsi - \bar{\eta}\,,
\\
f' ~=~ f - {h}\,, 
&\qquad 
\bar{f}' ~=~ \bar{f} - \bar{h}\,,
\end{aligned}
\end{equation}
which is in accordance with \eqref{gauge_transf_relations}.

\subsection{SUSY invariance}

The set of SUSY transformations of each field component can be derived by applying the generators of the SUSY algebra ($Q_{\alpha}$ and $\bar{Q}_{\dot{\alpha}}$) on the superfields, as a covariant transformation. Namely, the SUSY variation is given by
\begin{align}
\label{susy_variation}
\delta ~=~ \left( i\epsilon^{\alpha}Q_{\alpha} + i\bar{\epsilon}_{\dalpha}\bar{Q}^{\dalpha} \right)\,,
\end{align}
with the SUSY generators
\begin{equation}
\begin{aligned}
\label{susy_generators}
Q_{\alpha} &~=~ -i\frac{\p}{\p\theta^{\alpha}} - \sigma^{\mu}_{\alpha\dot{\alpha}}\btheta^{\dot{\alpha}}\p_{\mu}
\\
\bar{Q}_{\dot{\alpha}} &~=~ i\frac{\p}{\p\btheta^{\dot{\alpha}}} + \theta^{\alpha}\sigma^{\mu}_{\alpha\dot{\alpha}}\p_{\mu}\,,
\end{aligned}
\end{equation}
satisfying the SUSY algebra
\begin{align}
\label{algebra1}
[ \delta_2 ,\, \delta_1] ~=~ 
\left( i{\epsilon_1}{\sigma}^{\mu}\bepsilon_2 - i{\epsilon_2}{\sigma}^{\mu}\bepsilon_1 \right)\p_{\mu}\,.
\end{align}

Applying \eqref{susy_variation}, with \eqref{susy_generators}, on the superfields $V$, $S$ and $\bar{S}$, which are given in \eqref{V} and \eqref{chiral_sufields}, and aware of the covariant nature of the SUSY transformation, one can obtain, for the components of $V$, the following set of transformations,
\begin{equation}
\label{susy_transf1}
\begin{aligned}
\delta A_{\mu} &~=~ 
\epsilon\left(i\sigma_{\mu}\bar{\lambda}-\p_{\mu}\chi\right) 
+ \bar{\epsilon}\left(i\bar{\sigma}_{\mu}\lambda - \p_{\mu}\bar{\chi}\right)\,,
\\
\delta \bar{\lambda} 
&~=~ 
{2\bar{\sigma}^{\mu\nu}}\bar{\epsilon}\p_{\nu}A_{\mu} 
- i\bar{\epsilon}D 
~=~ 
\frac{1}{2}\bar{\sigma}^{\nu}\sigma^{\mu}\bar{\epsilon}F_{\mu \nu} 
- i\bar{\epsilon}D\,,
\\
\delta \lambda &~=~ 
{2\sigma^{\mu\nu}}\epsilon\p_{\nu}A_{\mu} 
+ i\epsilon D 
~=~ 
\frac{1}{2}\sigma^{\nu}\bar{\sigma}^{\mu}\epsilon F_{\mu \nu} 
+ i\epsilon D\,,
\\
\delta D &~=~ 
\bar{\epsilon}\bar{\sigma^{\mu}}\p_{\mu}\lambda 
- \epsilon\sigma^{\mu}\p_{\mu}\bar{\lambda}\,,
\\
\delta C &~=~ 
i\epsilon \chi 
- i \bar{\epsilon} \bar{\chi}\,,
\\
\delta \chi &=~ 
\epsilon M 
- i\sigma^{\mu} \bar{\epsilon} \left(A_{\mu} + i\p_{\mu}C\right)\,,
\\
\delta \bchi &~=~ 
\bar{\epsilon} \bar{M} 
- i\bar{\sigma}^{\mu} \epsilon \left(A_{\mu}-i\p_{\mu}C\right)\,,
\\
\delta M &~=~ 
2\bar{\epsilon}\left(\bar{\lambda} + i\bar{\sigma}^{\mu}\p_{\mu}\chi\right)\,,
\\
\delta \bar{M} &~=~ 
2\epsilon\left(\lambda + i\sigma^{\mu}\p_{\mu}\bar{\chi}\right)\,,
\end{aligned}
\end{equation}
and for the components of the Stueckelberg chiral and anti-chiral superfields, $S$ and $\bar{S}$, 
\begin{equation}
\label{susy_transf2}
\begin{aligned}
\delta \rho &~=~ 
\frac{\sqrt{2}}{2} \left( \epsilon \psi + \bar{\epsilon} \bar{\psi} \right)\,,
\\
\delta a &~=~ 
i\frac{\sqrt{2}}{2} \left( \bar{\epsilon} \bar{\psi} - \epsilon \psi \right)\,,
\\
\delta \psi &~=~ 
- \sqrt{2}{\epsilon}{f} 
+ i\sqrt{2}{\sigma}^{\mu}\bar{\epsilon} \p_{\mu}(\rho+i a)\,,
\\
\delta \bar{\psi} &~=~ 
- \sqrt{2}\bar{\epsilon}\bar{f} 
+ i\sqrt{2}\bar{\sigma}^{\mu}\epsilon \p_{\mu}(\rho-i a)\,,
\\
\delta f &~=~ 
- i\sqrt{2}  \bar{\epsilon}\bsigma^{\mu}\p_{\mu}\psi\,,
\\
\delta \bar{f} &~=~ 
- i \sqrt{2}\epsilon \sigma^{\mu}\p_{\mu}\bar{\psi}\,.
\end{aligned}
\end{equation}
Note that up to this point, no gauge fixing has been imposed and one may verify that the ASSM \eqref{action2} is explicit invariant under the SUSY transformations \eqref{susy_transf1} and \eqref{susy_transf2}, i.e.
\begin{align}
\label{invariance1}
\delta\left[\frac{1}{64}\int\, d^{4}x\, d^{2}\theta \ \ {W}(V){W}(V) + {c.c.} \right] ~=~ 0
\end{align}
and
\begin{equation}
\label{invariance2}
\delta\int\, d^{4}x\, d^{2}\theta \, d^{2}\btheta \ \ V^{H}V^{H} ~=~ 0\,.
\end{equation}

However, after imposing a gauge fixing condition, the supersymmetry transformation \eqref{susy_variation} must be changed in order compensate the gauge fixing, otherwise a gauge transformed superfield $V'$ would no longer satisfy the imposed gauge conditions \cite{Gates:1983nr,Buchbinder:1998twe}. In the section \ref{The Wess-Zumino gauge model}, we will explicit show that once the WZ gauge fixing condition is imposed, the original SUSY transformation \eqref{susy_variation} is no longer a symmetry of the action, due to the fact that $\delta V_{\rm wz}$ does not satisfy the WZ gauge conditions.

\section{The Wess-Zumino gauge}
\label{The Wess-Zumino gauge model}
The WZ gauge stands for a specific set of conditions on the field components of $V$ so that it does not depend on the compensating fields $(C,\, \chi,\, \bchi,\, M,\, \bar{M})$, but depends only on the field components $(A_{\mu},\, \lambda,\, \blambda,\, D)$. That is, the WZ gauge condition enforces the compensating fields to be null: $C_{\rm wz} = \chi_{\rm wz} = \bchi_{\rm wz} = M_{\rm wz} = \bar{M}_{\rm wz} = 0$ \cite{Wess:1974jb}. From \eqref{vec_field_gtransf}, one may verify that such a condition can be imposed by choosing,
\begin{equation}
\begin{aligned}
\label{wzg}
2 {\rm Im}(z) &~=~ 
C\,, 
\quad 
\sqrt{2}\eta ~=~ 
- \chi\,, 
\quad 
\sqrt{2}\bar{\eta} ~=~ 
- \bchi\,,
\\
2h &~=~ M\,, 
\quad \text{and} \quad 
2\bar{h} ~=~ \bar{M}\,.
\end{aligned}
\end{equation}
As a consequence, which can be verified with \eqref{chi_field_gtransf}, the (anti-)chiral's field components become,
\begin{equation}
\begin{aligned}
\label{wzg_chirals}
a_{\rm wz} ~=~ & 
a - \frac12 C\,, 
\\
\psi_{\rm wz} ~=~ 
\psi 
+ \frac{\sqrt{2}}{2} \chi \,, 
&\qquad 
\bpsi_{\rm wz} ~=~ 
\bpsi 
+ \frac{\sqrt{2}}{2}{\bchi}\,,
\\
f_{\rm wz} ~=~ 
f - \frac12 M\,, 
&\qquad 
\bar{f}_{\rm wz} ~=~ 
\bar{f} 
- \frac12 \bar{M}\,.
\end{aligned}
\end{equation}
Thus, instead of working with the original field components, $a,\, \psi,\, \bpsi,\, f$ and $\bar{f}$, we now have the variables in the WZ gauge, $a_{\rm wz}$, $\psi_{\rm wz}$, $\bpsi_{\rm wz}$, $f_{\rm wz}$ and $\bar{f}_{\rm wz}$, so that the Stueckelberg sector, and consequently $V^H_{\rm wz}$, do not explicitly depends on the compensating fields (see equations \eqref{Vwz}, \eqref{chiral_sufields_wz} and \eqref{wzVH} below). Notice that, otherwise, by working with the original field components of the Stueckelberg sector, $(a,\, \psi,\, \bpsi,\, f,\, \bar{f})$, the $V^{S}_{\rm wz}$ will have exactly the same expression as $V^{S}$, given by \eqref{VH}, due to its gauge invariance, and the mass term $\mathcal{M}(V^S)$ would still be given by \eqref{mass}, with an explicit dependence on $(C,\, \chi,\, \bchi,\, M,\, \bar{M})$.

After imposing the WZ gauge one can verify, from \eqref{vec_field_gtransf} and \eqref{chi_field_gtransf}, that the model still have a residual gauge freedom,
\begin{align}
\label{resid_gauge_free}
A'_{\mu} ~=~ A_{\mu} - 2 \p_{\mu}Re(z) \quad \text{and} \quad \rho' ~=~ \rho - Re(z)\,,
\end{align}
so that the longitudinal and transversal modes of $A_{\mu}$ still have to be accounted in the path integral quantization through the Faddeev-Popov procedure. Finally, in the WZ gauge the vector and the chiral superfields become
\begin{align}
\label{Vwz}
V_{\rm wz} &~=~ 
\theta \sigma^{\mu}\btheta A_{\mu} 
+ i\theta^{2}\btheta \blambda 
- i\btheta^{2}\theta \lambda
+ \frac12 \theta^{2}\btheta^{2}  D
\end{align}
and
\begin{equation}
\begin{aligned}
\label{chiral_sufields_wz}
S_{\rm wz} &~=~ (\rho + ia_{\rm wz}) + \sqrt{2} \theta \psi_{\rm wz} + i\theta\sigma^{\mu} \btheta\p_{\mu}(\rho + ia_{\rm wz}) -  \theta^2 f_{\rm wz} - \frac{i}{\sqrt{2}} \theta^2 \p_{\mu}\psi_{\rm wz} \sigma^{\mu}\btheta - \frac{1}{4} \theta^2 \btheta^2 \p^{2}(\rho + ia_{\rm wz})\,, 
\\
\bar{S}_{\rm wz}
&~=~ (\rho - ia_{\rm wz}) + \sqrt{2} \btheta \bpsi_{\rm wz} - i\theta\sigma^{\mu} \btheta\p_{\mu}(\rho - ia_{\rm wz}) -  \btheta^{2} \bar{f}_{\rm wz} + \frac{i}{\sqrt{2}} \btheta^2 \theta \sigma^{\mu}\p_{\mu}\bpsi_{\rm wz} - \frac{1}{4} \btheta^2 \theta^2 \p^{2}(\rho - ia_{\rm wz})\,,
\end{aligned}
\end{equation}
which is just like \eqref{chiral_sufields}, but with the subscript ``wz''. The $V^{S}_{\rm wz}$ then becomes,
\begin{align}
\label{wzVH}
V^{S}_{\rm wz} &~=~ 
- 2a_{\rm wz}
+ i\theta \sqrt{2}\psi_{\rm wz}
- i\btheta \sqrt{2}\bpsi_{\rm wz}
+ \theta\sigma^{\mu}\btheta \left[ A_{\mu} - 2\p_{\mu}\rho \right]
+ i \theta^{2}f_{\rm wz}
- i\btheta^{2}\bar{f}_{\rm wz}
\nonumber \\
& \phantom{~=~}
+ i\theta^{2}\btheta \left[ \blambda + \frac{i}{\sqrt{2}} \bsigma^{\mu}\p_{\mu} \psi_{\rm wz} \right]
- \btheta^{2}\theta \left[ \lambda + \frac{i}{\sqrt{2}}\sigma^{\mu} \p_{\mu} \bpsi_{\rm wz} \right]
+ \frac12 \theta^{2}\btheta^{2} \left[ D + \p^{2}a_{\rm wz} \right]\,.
\end{align}
The ASSM \eqref{action2} in the WZ gauge becomes,
\begin{align}
\label{wz_act}
\mathcal{S}_{\rm wz} &~=~
\int\, d^{4}x \ \left\{
- \frac{1}{4} f^{\mu\nu} f_{\mu\nu} 
- i \lambda \sigma^{\mu} \p_{\mu}\bar{\lambda}
+ \frac12 D^{2}
\right\}
+ m^2 \int\, d^{4}x\bigg\{
\frac{1}{2}{A^{S}}^{\mu}A_{\mu}^{S} 
- \sqrt{2}\bpsi_{\rm wz}\blambda 
- \sqrt{2}\psi_{\rm wz}\lambda 
\nonumber \\
& \phantom{\int\, d^{4}x \ }
- 2i\psi_{\rm wz}\sigma^{\mu}\p_{\mu}\bpsi_{\rm wz}
- 2\left(D + \p^{2}a_{\rm wz}\right) a_{\rm wz}
+2\bar{f}_{\rm wz}f_{\rm wz}\bigg\}\,,
\end{align}
with the gauge invariant $A^{S}_{\mu} ~=~ (A_{\mu} - 2\p_{\mu}\rho)$. The ASSM \eqref{wz_act} is the same, as it should, as the one studied in \cite{Buchbinder:1998twe}.

Now, let us check the SUSY invariance in the WZ gauge. The new set of SUSY transformation can be derived by applying the SUSY variation \eqref{susy_variation} on the vector superfield $V_{\rm wz}$, given in equation \eqref{Vwz}, and on the chiral superfields \eqref{chiral_sufields_wz}. Namely, the new set of SUSY transformation in the WZ gauge reads,
\begin{equation}
\label{wzsusytransf}
\begin{aligned}
\delta A_{\mu} &~=~ 
i\epsilon\sigma_{\mu}\bar{\lambda} 
+ i\bar{\epsilon}\bar{\sigma}_{\mu}\lambda
\\
\delta \bar{\lambda} &~=~ 
\frac{1}{2}\bar{\sigma}^{\nu}\sigma^{\mu}\bar{\epsilon}F_{\mu \nu} 
- i\bar{\epsilon}D
\\
\delta \lambda &~=~ 
\frac{1}{2}\sigma^{\nu}\bar{\sigma}^{\mu}\epsilon F_{\mu \nu} 
+ i\epsilon D
\\
\delta D &~=~ 
\bar{\epsilon}\bar{\sigma^{\mu}}\p_{\mu}\lambda 
- \epsilon\sigma^{\mu}\p_{\mu}\bar{\lambda}\,,
\end{aligned}
\end{equation}
and
\begin{equation}
\label{wz_susy_transf}
\begin{aligned}
\delta \rho &~=~ \frac{\sqrt{2}}{2} \left( \epsilon \psi_{\rm wz} 
+ \bar{\epsilon} \bar{\psi}_{\rm wz} \right)
\\
\delta a_{\rm wz} &~=~ i\frac{\sqrt{2}}{2} \left( \bar{\epsilon} \bar{\psi}_{\rm wz} - \epsilon \psi_{\rm wz} \right)
\\
\delta \psi_{\rm wz} &~=~ 
- \sqrt{2}{\epsilon}{f_{\rm wz}} 
+ i\sqrt{2}{\sigma}^{\mu}\bar{\epsilon} \p_{\mu}(\rho+i a_{\rm wz})
\\
\delta \bar{\psi}_{\rm wz} &~=~ 
- \sqrt{2}\bar{\epsilon}\bar{f}_{\rm wz} 
+ i\sqrt{2}\bar{\sigma}^{\mu}\epsilon \p_{\mu}(\rho-i a_{\rm wz})
\\
\delta f_{\rm wz} &~=~ 
- i\sqrt{2}\bar{\epsilon}\bsigma^{\mu}\p_{\mu}\psi_{\rm wz}
\\
\delta \bar{f}_{\rm wz} &~=~ 
- i \sqrt{2}\epsilon \sigma^{\mu}\p_{\mu}\bar{\psi}_{\rm wz}\,.
\end{aligned}
\end{equation}
Notice that the SUSY transformations of the field components from the chiral sector, \eqref{wz_susy_transf}, have exactly the same expression as those of \eqref{susy_transf2} (before fixing WZ gauge) but written in terms of the new WZ variables,

As one should expect, the dynamical sector of the action \eqref{wz_act} is invariant under \eqref{wzsusytransf}. That is,
\begin{align}
\delta \int\, d^{4}x \ \left\{
-\frac{1}{4} f^{\mu\nu} f_{\mu\nu} 
- i \lambda \sigma^{\mu} \p_{\mu}\bar{\lambda}
+ \frac12 D^{2}
\right\} ~=~ 0\,,
\end{align}
However, the Stueckelberg mass term of \eqref{wz_act} displays an explicit violation under \eqref{wzsusytransf} and \eqref{wz_susy_transf}. Namely, the breaking term reads
\begin{align}
\label{susy_breaking_term}
&\delta \int\, d^{4}x\, d^{2}\theta \, d^{2}\btheta \ \ V^{S}_{\rm wz}V^{S}_{\rm wz} ~=~ \nonumber\\
~=~&
\int\, d^{4}x \left\{
\epsilon \left[ 
i\sigma^{\mu}\blambda A_{\mu} 
- \frac{\sqrt{2}}{2}\sigma^{\mu}\bsigma^{\nu}\p_{\nu}\psi_{\rm wz} A_{\mu}
+ 2\lambda f_{\rm wz}
\right] 
\right.
\nonumber\\
& \phantom{\int\, d^{4}x}
\left.
+ \bar{\epsilon} \left[
i\bsigma^{\mu}\lambda A_{\mu} 
- \frac{\sqrt{2}}{2} \bsigma^{\mu}\sigma^{\nu}\p_{\nu}\bar{\psi}_{\rm wz}A_{\mu}
+ 2\blambda \bar{f}_{\rm wz}
\right] + \cdots
\right\} ~\neq~ 0\,.
\end{align}
This explicitly shows that our starting action in the WZ gauge, \eqref{wz_act}, is not invariant under the set of transformations \eqref{wzsusytransf} and \eqref{wz_susy_transf}. As we could see, prior to fixing the WZ gauge the ASSM \eqref{action2} is explicit SUSY invariant (see equations \eqref{invariance1} and \eqref{invariance2}). Actually, this is not a surprising fact, since it is well known that the once the gauge fixing condition is imposed the algebra of the original SUSY transformation \eqref{susy_variation} does not closes only on translation anymore (as in \eqref{algebra1}), but instead the algebra closes on translations plus gauge transformations, in order to compensate the gauge fixing ({\it cf.} text books \cite{Gates:1983nr,Buchbinder:1998twe} for details). In the Appendix \ref{appendix_B}, we show that it is possible to construct a proper gauge invariant action (no WZ gauge fixing) that is explicitly independent of the compensating fields, and show that, even in this case, the SUSY transformation does also have to take into account gauge fixing conditions, once the WZ gauge is imposed.

\section{The actual set of supersymmetry transformation}
\label{section4}

As verified in the previous section, once the WZ gauge is imposed, the original set of supersymmetry transformation obtained with \eqref{susy_variation} is no longer a symmetry of the ASSM. This is a side effect of considering a SUSY transformation \eqref{susy_variation} that does not satisfy the WZ gauge conditions when applied on a superfield $V_{\rm wz}$. As is well known from the literature, \cite{Gates:1983nr,Buchbinder:1998twe}, the correct set of SUSY transformation in the WZ gauge must take into account the gauge conditions,
\begin{align}
\label{susy_variation_wz}
\delta_{\rm wz} ~=~ \left( i\epsilon^{\alpha}Q_{\alpha} + i\bar{\epsilon}_{\alpha}\bar{Q}^{\alpha} \right)
+ i(\Lambda - \bar{\Lambda})_{\rm wz}\,,
\end{align}
where the (anti-)chiral sector \(i(\Lambda - \bar{\Lambda})_{\rm wz}\) is given by \eqref{gauge_tranf_param} with the WZ gauge conditions \eqref{wzg}. The existence of such new transformation relies on the SUSY algebra closing on translation plus gauge transformations,
\begin{align}
\left[ \delta_2 \,, \delta_1 \right] \varphi_{\rm wz}(x) ~=~ 
\left( i{\epsilon_1}{\sigma}^{\mu}\bepsilon_2 - i{\epsilon_2}{\sigma}^{\mu}\bepsilon_1 \right) \p_{\mu} \varphi_{\rm wz}(x) 
+ \delta_{\rm gauge}(\omega)\varphi_{\rm wz}(x)\,,
\label{susy_algebra1}
\end{align}
with
\begin{align}
\omega ~=~ \left( i{\epsilon_1}{\sigma}^{\mu}\bepsilon_2 - i{\epsilon_2}{\sigma}^{\mu}\bepsilon_1 \right)A_{\mu}(x)\,.
\end{align}
Applying the new transformation rule \eqref{susy_variation_wz} one gets the new set of SUSY transformations,
\begin{equation}
\label{newsym}
\begin{aligned}
\delta_{\rm wz} A_{\mu} &~=~ 
i\epsilon\sigma_{\mu}\bar{\lambda} 
+ i\bar{\epsilon}\bar{\sigma}_{\mu}\lambda
\\
\delta_{\rm wz} \bar{\lambda} &~=~ 
\frac{1}{2} \bar{\sigma}^{\nu} \sigma^{\mu} \bar{\epsilon} F_{\mu \nu} 
- i\bar{\epsilon} D
\\
\delta_{\rm wz} \lambda &~=~ 
	\frac{1}{2}\sigma^{\nu} \bar{\sigma}^{\mu} \epsilon F_{\mu \nu} 
+ i\epsilon D
\\
\delta_{\rm wz} D &~=~ 
- \epsilon\sigma^{\mu}\p_{\mu}\bar{\lambda}
+ \bar{\epsilon}\bar{\sigma^{\mu}}\p_{\mu}\lambda 
\\
\delta_{\rm wz} \rho &=~ 
\frac{\sqrt{2}}{2} \left(
\epsilon \psi_{\rm wz}
+ \bar{\epsilon} {\bpsi}_{\rm wz}
\right)
\\
\delta_{\rm wz} a_{\rm wz} &=~ 
i\frac{\sqrt{2}}{2} 
\left(\bar{\epsilon} {\bpsi}_{\rm wz}
- \epsilon \psi_{\rm wz} \right)
\\
\delta_{\rm wz} \psi_{\rm wz} &=~ 
- \sqrt{2}{\epsilon}f_{\rm wz} 
+ i\sqrt{2}{\sigma}^{\mu}\bar{\epsilon}\left[\p_{\mu}\rho + i\p_{\mu}a_{\rm wz} - \frac12 A_{\mu} \right] 
\\
\delta_{\rm wz} {\bpsi}_{\rm wz} &=~ 
-\sqrt{2}\bar{\epsilon}\bar{f}_{\rm wz}
+ i\sqrt{2}\bar{\sigma}^{\mu}\epsilon \left[ \p_{\mu}\rho - i\p_{\mu}a_{\rm wz} - \frac12 A_{\mu} \right] 
\\
\delta_{\rm wz} f &=~ 
- \bar{\epsilon}\bar{\lambda}
- i\sqrt{2}  \bar{\epsilon}\bsigma^{\mu}\p_{\mu}\psi_{\rm wz}
\\
\delta_{\rm wz} \bar{f}_{\rm wz} &=~ 
- \epsilon\lambda 
- i \sqrt{2}\epsilon \sigma^{\mu}\p_{\mu}\bar{\psi}_{\rm wz}\,,
\end{aligned}
\end{equation}
which could be verified, field per field, that it closes on the algebra \eqref{susy_algebra1}. Now, one may verify that the ASSM in the WZ gauge, \eqref{wz_act}, is explicit invariant under \eqref{newsym},
\begin{align}
\label{wz_act_inv}
\delta_{\rm wz}\mathcal{S}_{\rm wz} &~=~ 0\,.
\end{align}
Notice that the new SUSY transformation of $A_{\mu}$, $\lambda$, $\blambda$ and $D$ are exactly the same as those in \eqref{wzsusytransf}, so that the invariance of the kinetic term $W(V)W(V)$ is immediate,
\eqref{newsym},
\begin{align}
\label{wz_ww_inv}
\delta_{\rm wz} \int\, d^{4}x \ \left\{
- \frac{1}{4} f^{\mu\nu} f_{\mu\nu} 
- i \lambda \sigma^{\mu} \p_{\mu}\bar{\lambda}
+ \frac12 D^{2}
\right\}
~=~ 0
\end{align}
The new interesting point is that now the (anti-)chiral field components $\left\{ \psi_{\rm wz},\, \bpsi_{\rm wz},\, f_{\rm wz},\, \Bar{f}_{\rm wz} \right\}$ transform into field components of the gauge superfield $V$ (see the last four lines of \eqref{newsym}), and the super Stueckelberg mass term is invariant under $\delta_{\rm wz}$,
\begin{align}
\label{wz_VHVH_inv}
\delta_{\rm wz}\int\, d^{4}x\bigg\{
\frac{1}{2}{A^{S}}^{\mu}A_{\mu}^{S} 
- \sqrt{2}\bpsi_{\rm wz}\blambda 
- \sqrt{2}\psi_{\rm wz}\lambda 
- 2i\psi_{\rm wz}\sigma^{\mu}\p_{\mu}\bpsi_{\rm wz}
- 2\left(D + \p^{2}a_{\rm wz}\right) a_{\rm wz}
+2\bar{f}_{\rm wz}f_{\rm wz}\bigg\}
~=~ 0 \,.
\end{align}

\section{Conclusion}
\label{Comments and conclusions}

In this paper the matter of the invariance of the Abelian super Stueckelberg model (SSM) in the Wess-Zumino gauge, in the (3+1) Minkowski spacetime, under supersymmetric transformation is carefully revisited. The ASSM is considered here as a limit case of the super Yang-Mills Stueckelberg-like model, proposed in \cite{Capri:2018gpu}, with only the quadratic contribution of $V^{S} = V + i(S - \bar{S})$. In the equations \eqref{abelian_act} and \eqref{mass} the extended version of the ASSM, where the mass term is a power series of $V^{S}$, is written in the (3+1) dimensions Minkowski spacetime up to the 4th order, but in order to illustrate our point, only contributions up to $V^{S}V^{S}$ are needed (see the action \eqref{action2}).

Our analysis started with a detailed discussion on the gauge and supersymmetric transformations, verifying that the ASSM is invariant under \eqref{susy_variation} while no gauge fixing is performed. In the sequence we verified that, once the Wess-Zumino gauge is imposed the action \eqref{wz_act} is no longer invariant under the transformation \eqref{susy_variation}, pointing to the fact that $\delta V_{\rm wz}$ does not satisfy the WZ gauge conditions. In fact, the actual set of SUSY transformations must compensate the WZ gauge fixing by taking into account the gauge conditions, \eqref{susy_variation_wz}, so that the algebra closes on translation plus gauge transformation, \eqref{susy_algebra1}. As a consequence, we could derive the set of SUSY transformation of each field component of $V^{S}$ that is, indeed, an explicit symmetry of the ASSM in the WZ gauge. Since the super Stueckelberg sector is given by (anti-)chiral superfields, $i(S - \bar{S})$, with its particular gauge transformation rules, and that the WZ gauge conditions are imposed in order to eliminate the compensating fields from the vector superfield $V$, the new set of SUSY transformations \eqref{newsym} of the field components of the Stueckelberg sector is plagued by the fields $A_{\mu}$,  $\lambda$ and $\blambda$ from $V$. Thus, we see that the SUSY transformation of the chiral sector must compensate the gauge fixing conditions imposed on the vector superfield.

Such analysis becomes of great importance when the UV unitarity and stability of the model is considered. As is well known, the ASSM, in it self, is a free theory, and thus UV stable (also renormalizable when coupled to matter, \cite{Delbourgo:1975uf}). Furthermore, since the field components of the Stueckelberg superfield sector compensate the gauge fixing of the vector superfield components, the existence of a linear nilpotent BRST transformation of the ASSM ensures its unitarity. Besides, the linear Ward Identity of the equation of motion of the field component $a$, from the Stueckelberg sector, controls the existence of higher order terms as powers of $a$. However, nothing prohibits to consider the EASSM, \eqref{abelian_act} and \eqref{mass}, from the start, without imposing the transversality condition on $V^{S}$. In that extended scenario the UV stability is still a subject of investigation, and the existence of new vertices coupled to powers of $a(x)$ is certainly physically important given the dynamic of the field $a(x)$. In particular, we see that the mass term of the gauge field can be written as a power series of $a$, \( \frac{1}{2}A_{\mu}A_{\mu} \left( \sum_{n} c_{n}\, a^{n}(x) \right) \), so that the dynamic of $a(x)$ should be investigated in order to verify if the gauge field mass term gets suppressed or enhanced at a given energy scale.

\section*{Acknowledgments}
This study was financed in part by the Coordenação de Aperfeiçoamento de Pessoal de Nível Superior – Brazil (CAPES) - Finance Code 88887.950542/2024-00. This work was partially financed by the FAPERJ under the grant SEI-$260003/005833/2024$ - APQ1.

\begin{appendix}

\section{Notation and conventions on $N=1$ supersymmetry}
\label{appendix_A}

\begin{itemize}
\item Weyl spinors:
\end{itemize}
They are two components complex valued Grassmann fields,
\begin{eqnarray}
    \psi = \begin{pmatrix}
\psi_{1}\\
\psi_{2}
\end{pmatrix}\,.
\end{eqnarray}
The $\left( \frac12, 0 \right)$ representation of the Weyl spinor transforms under the element $M \in SL(2,\mathbb{C})$ as
\begin{eqnarray}
    \psi'_{\alpha} = M_{\alpha}^{\beta}\psi_{\beta}\,,
\end{eqnarray}
while the $\left( 0,\frac12 \right)$ spinor representation transforms with the complex conjugate $M^{\ast} \in SL(2,\mathbb{C})$ as
\begin{eqnarray}
    \bpsi'_{\dalpha} = (M^{*})^{\dbeta}_{\dalpha}\bpsi_{\dbeta}\,.
\end{eqnarray}
To lower and raise indices
\begin{eqnarray}\label{Indices}
    \psi^{\alpha} = \epsilon^{\alpha \beta}\psi_{\beta} \qquad \psi^{\dalpha} = \epsilon^{\dalpha \dbeta}\psi_{\dbeta} 
\end{eqnarray}
with
\begin{align}
    \epsilon^{\alpha \beta} = \epsilon^{\dalpha \dbeta} = \begin{pmatrix}
0&1\\
-1&0
\end{pmatrix} 
\quad \text{and} \quad
\epsilon_{\alpha \beta} = \epsilon_{\dalpha \dbeta} = 
\begin{pmatrix}
0&-1\\
1&0
\end{pmatrix}  \,.
\end{align}
The inner product between two spinor fields $\psi$ and $\chi$ (and their complex conjugate),
\begin{eqnarray}
\label{inner_prod}
    &&\psi\chi = \psi^{\alpha}\chi_{\alpha} = -\psi_{\alpha}\chi^{\alpha} = \chi^{\alpha}\psi_{\alpha} = -\chi_{\alpha}\psi^{\alpha} = \chi\psi
    \\
    &&\bpsi\bchi = \bpsi_{\dalpha}\bchi^{\dalpha} = - \bpsi^{\dalpha}\chi_{\dalpha} = \bchi_{\dalpha}\bpsi^{\dalpha} = -\bchi^{\dalpha}\bpsi_{\dalpha} = \bchi\bpsi\,,
\end{eqnarray}
with Greek indices $\alpha,\,\beta,\,\gamma,\,\delta = 1,\,2$.

\begin{itemize}
    \item The Pauli matrices:
\end{itemize}
\begin{align*}
&&\sigma_{1} =\begin{pmatrix}
0&1\\
1&0
\end{pmatrix}, &&\sigma_{2} =\begin{pmatrix}
0&-i\\
i&0
\end{pmatrix}, &&\sigma_{3}=\begin{pmatrix}
1&0\\
0&-1
\end{pmatrix}\,,
\end{align*}
with 
$
    \Tr( \sigma^{\mu} \bsigma^{\nu}) = 2\eta^{\mu \nu}\,.
$
\begin{itemize}
\item The spacetime metric: $\delta_{\mu\nu}=diag(1,\,-1,\,-1,\,-1)$, with Greek indices ($\mu,\,\nu,\, \lambda,\, \rho,\,\tau ~=~ 0,\,1,\,2,\, 3$).
\item Useful identities:
\end{itemize}
\begin{eqnarray}
    && \Big(\sigma^{\mu}\bsigma^{\nu} + \sigma^{\nu}\bsigma^{\mu} \Big)^\beta_\alpha = 2g^{\mu \nu}\delta_{\dalpha}^{\beta}
    \\
    &&\Big(\bsigma^{\mu}\sigma^{\nu} + \bsigma^{\nu}\sigma^{\mu}\Big)_{\dbeta}^{\dalpha} = 2g^{\mu \nu}\delta_{\dbeta}^{\dalpha}
\end{eqnarray}
\begin{eqnarray}
    &&\sigma^{\mu \nu \beta}_{\alpha} = \frac{1}{4}\Big(\sigma^{\mu}_{\alpha \dalpha}\bsigma^{\nu \dalpha \beta} - \sigma^{\nu}_{\alpha\dalpha}\bsigma^{\mu \dalpha \beta}\Big)
    \\
    &&\bsigma^{\mu \nu \dalpha}_{\dbeta} = \frac{1}{4}\Big(\bsigma^{\mu \dalpha \alpha}\sigma^{\nu}_{\alpha \dbeta} - \bsigma^{\nu \dalpha \alpha}\sigma^{\mu}_{\alpha \dbeta}\Big) 
\end{eqnarray}
\begin{eqnarray}
   && \sigma^{\mu}_{\alpha \dalpha}\sigma_{\mu}^{\beta \dbeta} = 2\delta^{\beta}_{\alpha} \delta^{\dbeta}_{\dalpha}, \quad \ \qquad \qquad \qquad \sigma^{\mu}_{\alpha \dalpha}\sigma_{\mu \beta \dbeta} = 2\varepsilon_{\alpha \beta} \varepsilon_{\dalpha \dbeta},
   \\
   &&\sigma_{\mu \alpha \dalpha} \sigma^{\mu \nu}_{\beta \gamma} = i(\varepsilon_{\alpha \beta}\sigma^{\nu}_{\gamma \alpha}+\varepsilon_{\alpha \gamma}\sigma^{\nu}_{\beta \dalpha}), \quad \sigma_{\mu \alpha\dalpha}\overline{\sigma}^{\mu \nu}_{\beta \dot{\gamma}} = -i(\varepsilon_{\dalpha \dbeta}\sigma^{\nu}_{\alpha \dot{\gamma}} + \varepsilon_{\dalpha \dot{\gamma}}\sigma^{\nu}_{\alpha \dbeta}),
   \\
   &&\sigma^{\alpha\beta}_{\mu \nu}\sigma^{\mu \nu}_{\gamma \delta} = -4(\delta^{\alpha}_{\gamma}\delta^{\beta}_{\delta}+\delta^{\alpha}_{\delta}\delta^{\beta}_\gamma), \quad \quad \overline{\sigma}^{\dalpha \dbeta}_{\mu \nu}\overline{\sigma}^{\mu \nu}_{\dot{\gamma} \dot{\delta}} = -4(\delta^{\dalpha}_{\dot{\gamma}} \delta^{\dbeta}_{\dot{\delta}} + \delta^{\dalpha}_{\dot{\gamma}}\delta^{\dbeta}_{\dot{\delta}}),
   \\
   &&\sigma^{\dalpha \dbeta}_{\mu \nu} \sigma^{\mu \nu}_{\gamma \delta} = 0.
\end{eqnarray}
\begin{eqnarray}
    &&\frac{1}{2}\varepsilon^{\mu \nu \rho \sigma}\sigma_{\rho \sigma} = - i\sigma^{\mu \nu}, \quad \frac{1}{2}\varepsilon^{\mu \nu \rho \sigma} \overline{\sigma}_{\rho \sigma} = i\overline{\sigma}^{\mu \nu}
    \\
    &&\varepsilon_{\mu \nu \rho}^{\tau}\sigma_{\tau \lambda} = i\sigma_{\mu \nu}g_{\rho \lambda} - \sigma_{\mu \rho}g_{\nu \lambda}+i\sigma_{\nu \rho}g_{\mu \lambda},
    \\
    &&\varepsilon_{\mu \nu \rho}^{\tau}\overline{\sigma}_{\tau \lambda} = -i\overline{\sigma}_{\mu \nu}g_{\rho \lambda} + \overline{\sigma}_{\mu \rho}g_{\nu \lambda}-i\overline{\sigma}_{\nu \rho}g_{\mu \lambda}
\end{eqnarray}
with $\varepsilon_{0123} = 1 = -\varepsilon^{0123}$.
\begin{eqnarray}
    \sigma_{\mu} \overline{\sigma}_{\nu} = g_{\mu \nu} - i\sigma_{\mu \nu}, \quad \overline{\sigma}_{\mu}\sigma_{\nu} = g_{\mu \nu} - i\overline{\sigma}_{\mu \nu}.
\end{eqnarray}
\begin{eqnarray}
    &&\sigma^{\mu}\overline{\sigma}^{\nu}\sigma^{\rho} = g^{\mu \nu}\sigma^{\rho} + g^{\nu \rho}\sigma^{\mu} - g^{\mu \rho}\sigma^{\nu} -i\varepsilon^{\mu \nu \rho \lambda}\sigma_{\lambda}.
    \\
  &&\overline{\sigma}^{\mu}\sigma^{\nu}\overline{\sigma}^{\rho} = g^{\mu \nu}\overline{\sigma}^{\rho} + g^{\nu \rho}\overline{\sigma}^{\mu} 
  - g^{\mu \rho}\bsigma^{\nu}
  + i\varepsilon^{\mu \nu \rho \lambda}\overline{\sigma}_{\lambda}.
\end{eqnarray}

\begin{itemize}
\item The superspace:
\end{itemize}
It can be defined as an eight dimension space with the four spacetime $x^{\mu}$ and four Grassmann valued spinorial coordinates, $(\theta^{\alpha},\btheta^{\dalpha})$.
The inner product between two spinorial coordinates of superspace is given by the same rules of inner products between generic spinors \eqref{inner_prod}. Here follow some useful identities with the Grassmann coordinates,
\begin{eqnarray}
    &&\theta^{\alpha}\theta^{\beta} = - \frac{1}{2}\epsilon^{\alpha \beta} \theta \theta \quad, \quad \btheta^{\dalpha} \btheta^{\dbeta} = \frac{1}{2}\epsilon^{\dalpha \dbeta}\btheta \btheta,
    \\
    &&\theta_{\alpha}\theta_{\beta} = \frac{1}{2} \epsilon_{\alpha \beta} \theta \theta \quad , \quad \btheta_{\dalpha} \btheta_{\dbeta} = -\frac{1}{2}\epsilon_{\dalpha \dbeta} \btheta \btheta, 
    \\
    &&\theta\sigma^{\mu}\btheta \theta\sigma^{\nu}\btheta = \frac{1}{2}\theta \theta \btheta \btheta g^{\mu \nu} \quad , \quad \theta \psi \theta \chi = -\frac{1}{2} \theta \theta \psi \chi
\end{eqnarray}
The derivatives
\begin{eqnarray}
    &&
    \frac{\partial\theta^{\beta}}{\partial \theta^{\alpha} } = \delta^{\beta}_{\alpha}\,, \quad
    \frac{\partial\btheta^{\dbeta}}{\partial \btheta^{\dalpha}} = \delta^{\dbeta}_{\dalpha}
    \quad \text{and} \quad
    \frac{\partial\btheta^{\dbeta}}{\partial \theta^{\alpha} } = 0\,, \\
    &&
    \frac{\partial\theta_{\beta}}{\partial \theta^{\alpha} } = \epsilon_{\beta \alpha}\,,
    \frac{\partial\btheta_{\dbeta}}{\partial \btheta^{\dalpha}} = \epsilon_{\dbeta \dalpha}\,.
\end{eqnarray}

\begin{itemize}
\item Chiral superfields:
\end{itemize}
Chiral and anti-chiral superfields are defined, respectivelly, by the conditions
\begin{align}
\bar{D}_{\alpha} \Lambda ~=~ 0
\quad \text{and} \quad
{D}_{\alpha} \bar{\Lambda} ~=~ 0 \,.
\end{align}
with the covariant derivatives,
\begin{align}
D_{\alpha} ~=~ \frac{\p}{\p\theta^{\alpha}} + i\sigma^{\mu}_{\alpha\dalpha}\btheta^{\dalpha}\p_{\mu}
\quad \text{and} \quad
\bar{D}_{\dalpha} ~=~ \frac{\p}{\p\btheta^{\dalpha}} 
+ i\theta^{\alpha}\sigma^{\mu}_{\alpha\dalpha}\p_{\mu}
\end{align}

\section{An action free from compensating fields}
\label{appendix_B}

In this section we propose an Abelian Stueckelberg-like model (SSM) that is explicitly free from the compensating fields and is manifestly gauge invariant. In other words, our proposed action have the same expression as the one in the WZ gauge, \eqref{wz_act}, but without imposing any gauge condition. Since there will be no gauge fixing, the vector superfield $V$ remains intact, as given by equation \eqref{V}. In the chiral sector, however, we perform a shift in the following field components,
\begin{equation}
	\label{shift}
\begin{aligned}
&a ~\to~ a + \frac{1}{2}C\,, \\
&\psi ~\to~ \psi - \frac{\sqrt{2}}{2}\chi\,, \quad
\bpsi ~\to~ \bpsi - \frac{\sqrt{2}}{2}\bchi\,, \\
&f ~\to~ f + \frac{1}{2}M\,, \quad
\bar{f} ~\to~ \bar{f} + \frac{1}{2}\bar{M}\,,
\end{aligned}
\end{equation}
so that the chiral and anti-chiral superfields become,
\begin{align}
\label{shifted_chiral}
S&~=~ 
\rho + i\left(a + \frac{C}{2}\right) 
+ \sqrt{2} \theta \left(\psi -\frac{\chi}{\sqrt{2}}\right) 
+ i\theta\sigma^{\mu} \btheta\p_{\mu}\left[\rho + \left(a + \frac{C}{2}\right)\right] 
-  \theta^2 \left(f +\frac{M}{2}\right) 
\nonumber \\
& \phantom{~=~}
+ \frac{i}{\sqrt{2}} \theta^2\btheta \bsigma^{\mu} \p_{\mu}\left(\psi -\frac{\chi}{\sqrt{2}}\right) 
+ \frac{1}{4} \theta^2 \btheta^2 \p^{2}\left[\rho + i\left(a + \frac{C}{2}\right)\right]\,, 
\\
\bar{S} &~=~ 
\label{shifted_antchiral}
\rho - i\left(a + \frac{C}{2}\right) 
+ \sqrt{2} \btheta \left(\bpsi -\frac{\bchi}{\sqrt{2}}\right) 
- i\theta\sigma^{\mu} \btheta\p_{\mu}\left[\rho - i\left(a + \frac{C}{2}\right)\right] 
-  \btheta^{2} \left(\bar{f}+\frac{\bar{M}}{2}\right) 
\nonumber \\
&\phantom{~=~}
+ \frac{i}{\sqrt{2}} \btheta^2 \theta \sigma^{\mu}\p_{\mu}\left(\bpsi -\frac{\bchi}{\sqrt{2}}\right)
-\frac{1}{4} \btheta^2 \theta^2 \p^{2}\left[\rho - i\left(a + \frac{C}{2}\right)\right]\,.
\end{align}
As a consequence, in terms of the new shifted variables, the $V^{S}$ becomes independent of the compensating fields,	
\begin{align}
\label{shifted_VH}
V^{S} &=~ V + i(S - \bar{S})
\nonumber \\
&=~
- 2a
+ i\theta \sqrt{2}\psi
- i\btheta \sqrt{2}\bpsi
+ \theta\sigma^{\mu}\btheta \big[ A_{\mu} - 2\p_{\mu}\rho \big]
+ i \theta^{2}f
- i\btheta^{2}\bar{f}
\nonumber \\
& \phantom{=~}
+ i\theta^{2}\btheta \left[ \blambda + \frac{i}{\sqrt{2}} \bsigma^{\mu}\p_{\mu} \psi \right]
- \btheta^{2}\theta \left[ \lambda + \frac{i}{\sqrt{2}}\sigma^{\mu} \p_{\mu} \bpsi \right]
+ \frac12 \theta^{2}\btheta^{2} \left[ D + \p^{2}a \right]\,,
\end{align}
just as in the WZ gauge \eqref{wzVH}. It should be stressed out that the new shifted variables are gauge invariant, which can be verified by applying the set of gauge transformations \eqref{vec_field_gtransf} and \eqref{chi_field_gtransf} on \eqref{shift}. Therefore, the gauge invariance feature of $V^{S}$ is preserved. Of course, since the field components of $V$ are not shifted, their gauge transformation rule \eqref{vec_field_gtransf} remains unchanged. As a consequence of the shift \eqref{shift}, our proposed Abelian Stueckelberg-like action, in its field component expression, becomes gauge invariant and explicitly independent of the compensating fields. Namely, the new Abelian Stueckelberg-like action reads
\begin{align}
\label{shift_act}
\mathcal{S}^{(2)} &=~
\int\, d^{4}x \ \left\{
-\frac{1}{4} f^{\mu\nu} f_{\mu\nu} 
- i \lambda \sigma^{\mu} \p_{\mu}\bar{\lambda}
+ \frac12 D^{2}
\right\}
\nonumber \\
& + m^2 \int\, d^{4}x\Bigg[
\frac{1}{2}{A^{S}}^{\mu}A_{\mu}^{S} 
- \sqrt{2}\bpsi\blambda 
- \sqrt{2}\psi\lambda - 2i\psi\sigma^{\mu}\p_{\mu}\bpsi
- 2\Bigg(D + \p^{2}a\Bigg) a
+2\bar{f}f\Bigg]\,,
\end{align}
with $A^{S}_{\mu} ~=~ (A_{\mu} - 2\p_{\mu}\rho)$. Notice that the shifted action \eqref{shift_act} reads just like the action in the WZ gauge (see equation \eqref{wz_act}), even though no gauge fixing condition has been imposed. So here, in the Abelian case, the WZ gauge fixing is unnecessary.

Let us now discuss the invariance of \eqref{shift_act} under the SUSY variation \eqref{susy_variation}. Since the superfield $V$ remains unchanged, the SUSY transformation of its field components are given by the ones in equation (\ref{susy_transf1}). On the other hand, the field components of the chiral sector have a slightly different set of transformation. Namely,
\begin{equation}
\label{shift_susy_transf2}
\begin{aligned}
\delta \rho &=~ 
\frac{\sqrt{2}}{2}(\epsilon \psi +   \bar{\epsilon} \bar{\psi}) 
- \frac{1}{2}(\epsilon \chi+\bar{\epsilon}\bchi) \,, 
\\
\delta a &=~ 
i\frac{\sqrt{2}}{2}(\bar{\epsilon} \bar{\psi} - \epsilon \psi) \,, 
\\
\delta \psi &=~ 
- \sqrt{2}{\epsilon}{f} 
- i\frac{\sqrt{2}}{2}{\sigma}^{\mu}\bar{\epsilon}\left[A^{S}_{\mu} - 2i\p_{\mu}a\right] \,, 
\\
\delta \bar{\psi} &=~ 
-\sqrt{2}\bar{\epsilon}\bar{f} 
- i\frac{\sqrt{2}}{2}\bar{\sigma}^{\mu}\epsilon \left[A^{S}_{\mu} + 2i\p_{\mu}a\right] \,,
\\
\delta f &=~ 
- \bar{\epsilon}\bar{\lambda}
- i\sqrt{2}  \bar{\epsilon}\bsigma^{\mu}\p_{\mu}\psi \,,
\\
\delta \bar{f} &=~ 
- \epsilon\lambda 
- i \sqrt{2}\epsilon \sigma^{\mu}\p_{\mu}\bar{\psi}\,.
\end{aligned}
\end{equation}
Notice that as an aftermath of the shift, the SUSY transformation of $\psi$ and $\bpsi$ depends on $A^{S}_{\mu}$, in order to compensate the dependence of $\rho$ on $\chi$ which is a field component of the vector superfield $V$. This is not something to worry about, since the proposed shift does mix up the field components of both sectors, the vector and chiral superfields.

Applying the set of SUSY transformations \eqref{susy_transf1} and \eqref{shift_susy_transf2} on the new action \eqref{shift_act} one may verify that $\delta S^{(2)} = 0$:
\begin{align}
	\label{shift_invariance1}
	\delta\left[\frac{1}{64}\int\, d^{4}x\, d^{2}\theta \ \ {W}(V){W}(V) + {c.c.} \right] ~=~ 0
\end{align}
and
\begin{equation}
	\label{shift_invariance2}
	\delta\int\, d^{4}x\, d^{2}\theta \, d^{2}\btheta \ \ V^{S}V^{S} ~=~ 0\,.
\end{equation}
The SUSY invariance of the dynamic term, \eqref{shift_invariance1}, is clear, since the shift \eqref{shift} is performed only in the chiral Stueckelberg-like sector and no gauge fixing is enforced. So the invariance \eqref{invariance1} is exactly the same as \eqref{shift_invariance1}. However, the interesting outcome is the invariance of the mass contribution \eqref{shift_invariance2}. It should be clear that the action \eqref{shift_act}, which is independent of the compensating fields, was written without imposing the WZ gauge condition and, as a consequence, the Stueckelberg-like contribution becomes SUSY invariant. That is, the WZ gauge is not needed in order to write down an ASSM that is independent of compensating fields and manifestly SUSY invariant.

Nevertheless, we must note that if we try to impose the WZ gauge condition \eqref{wzg} in this shifted scenario,
we will conclude that the resulting action will no longer be invariant by the set of transformations \eqref{susy_transf1} and \eqref{shift_susy_transf2}, as the SUSY transformation rule is sensible to the gauge fixing.

\end{appendix}

\end{document}